\documentclass[11pt]{article}
\usepackage{fullpage}
\usepackage{fancyhdr}

\usepackage{color}
\usepackage[cmex10]{amsmath} 
\usepackage{amssymb}
\usepackage{bbm} 
\usepackage{graphicx} 
\usepackage[caption=false,font=footnotesize]{subfig} 
\usepackage[rightcaption]{sidecap} 
\usepackage{floatrow}
\usepackage{tabu}
\newcolumntype{Y}{D..{2}}
\usepackage{dcolumn} 
\usepackage{booktabs}
\usepackage{fixltx2e} 
\usepackage{engord} 
\usepackage{xifthen} 
\usepackage{amsthm} 
\usepackage{nth}
\usepackage{changepage}

\usepackage[linesnumbered,ruled,nofillcomment]{algorithm2e}
\SetKwBlock{KwFunction}{function}{}
\SetKwBlock{KwOn}{on}{}
\SetKwBlock{KwPeriodically}{periodically}{}
\SetKwBlock{KwState}{state}{}
\SetKwBlock{KwAtomic}{atomically}{}
\SetKwBlock{KwConcurrently}{concurrently}{}
\SetCommentSty{textup} 
\SetKwComment{KwIttayComment}{(}{)} 
\SetKw{KwAnd}{and} 
\SetKw{KwOr}{or}
\SetKw{KwSetTimer}{setTimer}
\SetKwBlock{KwExpTimer}{on timer expiration}{}
\SetKwFor{ForAll}{forall}{}

\let\oldnl\nl
\newcommand{\nonl}{\renewcommand{\nl}{\let\nl\oldnl}}

\ifx\pdftexversion\undefined
\usepackage[colorlinks,linkcolor=black,filecolor=black,citecolor=black,urlcolor=black,pdfstartview=FitH]{hyperref}
\else
\usepackage[colorlinks,linkcolor=blue,filecolor=blue,citecolor=blue,urlcolor=blue,pdfstartview=FitH]{hyperref}
\fi

\usepackage{marginnote}

\newcommand{\bitcoinNG}{Bitcoin-NG}

\date{}

\title{ 
\Large \bf 
Bitcoin-NG: A Scalable Blockchain Protocol 
} 

\author{{\rm Ittay Eyal}\qquad {\rm Adem Efe Gencer}\qquad {\rm Emin G\"{u}n Sirer}\qquad  {\rm Robbert van Renesse} \\ Cornell University}

\begin{document} 

\maketitle 

\pagestyle{empty} 



\abstract{
Cryptocurrencies, based on and led by Bitcoin, have shown promise as infrastructure for pseudonymous online payments, cheap remittance, trustless digital asset exchange, and smart contracts. 
However, Bitcoin-derived blockchain protocols have inherent scalability limits that trade-off between throughput and latency and withhold the realization of this potential. 

This paper presents \bitcoinNG, a new blockchain protocol designed to scale.
Based on Bitcoin's blockchain protocol, \bitcoinNG\ is Byzantine fault tolerant, is robust to extreme churn, and shares the same trust model obviating qualitative changes to the ecosystem.

In addition to \bitcoinNG, we introduce several novel metrics of interest in quantifying the security and efficiency of Bitcoin-like blockchain protocols. 
We implement \bitcoinNG\ and perform large-scale experiments at~$15\%$ the size of the operational Bitcoin system, using unchanged clients of both protocols.
These experiments demonstrate that \bitcoinNG\ scales optimally, with bandwidth limited only by the capacity of the individual nodes and latency limited only by the propagation time of the network. 
} 
\newcommand{\secref}[1]{{\S}\ref{#1}}

    \section{Introduction} 

Bitcoin has emerged as the first widely-deployed, decentralized global currency, and sparked hundreds of copycat currencies. 
Overall, cryptocurrencies have garnered much attention from the financial and tech sectors, as well as academics, 
achieved wide market penetration in underground economies~\cite{meiklejohn2013fistful},  
reached a \$12B market cap and attracted close to \$1B in venture capital~\cite{coindesk2015capital}. 
The core technological innovation powering these systems is the \emph{Nakamoto consensus} protocol for maintaining a distributed ledger known as the blockchain. 
The blockchain technology provides a decentralized, open, Byzantine fault-tolerant transaction mechanism, and 
promises to become the infrastructure for a new generation of Internet interaction, including anonymous online payments~\cite{ashleyMadison2015}, remittance, and transaction of digital assets~\cite{coloredCoins2015}. Ongoing work explores smart digital contracts, enabling anonymous parties to programmatically enforce complex agreements~\cite{kosba2015hawk, ethereum2015white}. 
 
Despite its potential, blockchain protocols face a significant scalability barrier~\cite{sompolinsky2015ghost, lewenberg2015inclusive, decker2015duplex, bamert2013snack}. 
The maximum rate at which these systems can process transactions is capped by the choice of two parameters: block size and block interval. 
Increasing block size improves throughput, but the resulting bigger blocks take longer to propagate in the network. Reducing the block interval reduces latency, but leads to instability where the system is in disagreement and the blockchain is subject to reorganization. 
To improve efficiency, one has to trade off throughput for latency.
Bitcoin currently targets a conservative~10 minutes between blocks, yielding~10 minute expected latencies for transactions to be encoded in the blockchain.\footnote{On average, assuming no backlog, both block interval and the average time to wait for a block starting at any time are ten minutes. This is a non-intuitive property of the memoryless exponential distribution.} The block size is currently set at~1MB, yielding only~1 to~3.5 transactions per second for Bitcoin for typical transaction sizes.
Proposals for increasing the block size are the topic of heated debate within the Bitcoin community~\cite{peck2015coup}.

In this paper, we present \bitcoinNG, a scalable blockchain protocol, based on the same trust model as Bitcoin. 
\bitcoinNG's latency is limited only by the propagation delay of the network, and its bandwidth is limited only by the processing capacity of the individual nodes. 
{\bitcoinNG} achieves this performance improvement by decoupling Bitcoin's blockchain operation into two planes: \emph{leader election} and \emph{transaction serialization}. 
It divides time into epochs, where each epoch has a single leader. 
As in Bitcoin, leader election is performed randomly and infrequently.
Once a leader is chosen, it is entitled to serialize transactions unilaterally until a new leader is chosen, marking the end of the former's epoch. 

While this approach is a significant departure from Bitcoin's operation, \bitcoinNG\ maintains Bitcoin's security properties. 
Implicitly, leader election is already taking place in Bitcoin. But in Bitcoin, the leader is in charge of serializing history, making the entire duration of time between leader elections a long system freeze. 
In contrast, leader election in \bitcoinNG\ is forward-looking, and ensures that the system is able to continually process transactions.

Evaluating the performance and functionality of new consensus protocols is a challenging task. 
To help perform this quantitatively and provide a foundation for the comparison of alternative consensus protocols, we introduce several metrics to evaluate implementations of the Nakamoto consensus. 
These metrics capture performance metrics such as protocol goodput and latency, as well as various aspects of its security, including its ability to maintain consensus and resist centralization. 

We evaluated the performance of \bitcoinNG\ on a large emulation testbed consisting of~$1000$ nodes, 
amounting to over~$15\%$ of the current operational Bitcoin network~\cite{miller2015topology}. 
This testbed enables us to run unchanged clients, using realistic Internet latencies. 
We compare~\bitcoinNG\ with the original Bitcoin client, and demonstrate the critical tradeoffs inherent in the original Bitcoin protocol. 
Controlling for network bandwidth, reducing Bitcoin's latency by decreasing the block interval and improving its throughput by increasing the block size both yield adverse effects. 
In particular, fairness suffers, 
giving large miners an advantage over small miners. This anomaly leads to centralization, where the mining power tends to be used under a single controller, breaking the basic premise of the decentralized cryptocurrency vision. 
Additionally, mining power is lost, making the system more vulnerable to attacks. 
In contrast, \bitcoinNG\ improves latency and throughput to the maximum allowed by network conditions and node processing limits, while avoiding the fairness and mining power utilization problems. 

In summary, this paper makes three contributions. 
First, it outlines the \bitcoinNG\ scalable blockchain protocol, which achieves significantly higher throughput and lower latency than Bitcoin while maintaining the Bitcoin trust assumptions. 
Second, it introduces quantitative metrics for evaluating Nakamoto consensus protocols. 
These metrics are designed to ground the ongoing discussion over parameter selection in Bitcoin-derived currency.
Finally, it quantifies, through large-scale experiments, \bitcoinNG's robustness and scalability.

    \section{Model and Goal} \label{sec:model} 

\newcommand{\nodes}{\ensuremath{\mathcal{N}}} 

The system is comprised of a set of nodes $\nodes$
connected by a reliable authenticated peer-to-peer network. 
Each node can poll a random oracle~\cite{bellare1993oracle} as a random bit source. 
Nodes can generate key-pairs, but there is no trusted public key infrastructure. 

The system employs an associated puzzle system, defined by a cryptographic hash function~$H$. 
The solution to a puzzle defined by the string~$y$ is a string~$x$ such that $H(y|x)$~--- the hash of the concatenation of the two~--- is smaller than some target. 
Each node~$i$ has a limited amount of compute power, called \emph{mining power}, measured by the number of potential puzzle solutions it can try per second. 
A solution to a puzzle constitutes a \emph{proof of work}, as it statistically indicates the amount of work a node had to perform in order to find it. 

At any time~$t$, a subset of nodes $B(t) \subset \nodes$ are Byzantine and behave arbitrarily, controlled by a single adversary. 
The other nodes are \emph{honest}~--- they abide by the protocol. 
The mining power of each node~$i$ is~$m(i)$. 
The mining power of the Byzantine nodes is less than~1/4 of the total compute power at any given time: 

\[ 
\forall t: \sum_{b \in B(t)} m(b) < \frac{1}{4} \sum_{n \in \nodes} m(n)
\] 

\noindent
because proof-of-work blockchains, \bitcoinNG\ included, are vulnerable to selfish mining by attackers larger than~$1/4$ of the network~\cite{eyal2014majority}. 

        \subsection*{Nakamoto Consensus} 

The nodes are to implement a replicated state machine (RSM)~\cite{lamport1984timeout,schneider1990fault}. 
Properties of the system can be compared to those of classical consensus~\cite{pease1980agreement}: 

\begin{description} 
\itemsep0em

\item[Termination] There exists a time difference function $\Delta(\cdot)$ such that, given a time $t$ and a value $0 < \varepsilon < 1$, the probability is smaller than~$\varepsilon$ that at times $t', t'' > t + \Delta(\varepsilon)$ a node returns two different states for the machine at time~$t$. 

\item[Agreement] There exists a time difference function $\Delta(\cdot)$ such that, given a $0 < \varepsilon < 1$, the probability that at time $t$ two nodes returns different states for $t - \Delta(\varepsilon)$ is smaller than $\varepsilon$. 

\item[Validity] If the fraction of mining power of Byzantine nodes is bounded by $f$, 
$
\forall t: 
\frac{
    \sum_{b \in B(t)} m(b)
}{
    \sum_{n \in \nodes} m(n)
}
< f \,\,\, , 
$
then the average fraction of state machine transitions that are not inputs of honest nodes is smaller than $f$. 

\end{description}

    \section{Bitcoin and its Block\-chain Protocol} \label{sec:bitcoin} 
    
Bitcoin is a distributed, decentralized crypto-currency~\cite{bitcoin2015source, bitcoin2013rules, bitcoin2013protocol, nakamoto2008bitcoin}, which implicitly defined and implemented
the Nakamoto consensus. 
Bitcoin uses the blockchain protocol to serialize transactions of the Bitcoin currency among its users. 
The replicated state machine maintains the balance of the different users, and its transitions are transactions that move funds among them. This state machine is managed by the system nodes, called miners. 

Each user commands \emph{addresses}, and sends Bitcoins by forming a transaction from her address to another's address and sending it to the nodes. 
More explicitly, a transaction is from the output of a previous transaction, to a specific address. 
An output is \emph{spent} if it is the input of another transaction. 
A client owns~$x$ Bitcoins at time~$t$ if the aggregate of unspent outputs to its address is~$x$.
Transactions are protected with cryptographic techniques that ensure only the rightful owner of a Bitcoin address can transfer funds from it. 
Miners accept transactions only if their sources have not been spent, thereby preventing users from double-spending their funds. 
The miners commit the transactions into a global append-only log called the \emph{blockchain}. 

The blockchain records transactions in units of blocks. 
Each block includes a unique ID, and the ID of the preceding block. 
The first block, dubbed \emph{the genesis block}, is defined as part of the protocol. 
A valid block contains 
(1) a solution to a cryptopuzzle involving the hash of the previous block, 
(2) the hash (specifically, the Merkle root) of the transactions in the current block, which have to be valid, and 
(3) a special transaction, called the \emph{coinbase}, crediting the miner with the reward for solving the cryptopuzzle. 
This process is called Bitcoin \emph{mining}, and, by slight abuse of terminology, we refer to the creation of blocks as \emph{block mining}. 
The specific cryptopuzzle is a double-hash of the block header whose result has to be smaller than a set value. The \emph{problem difficulty}, set by this value, is dynamically adjusted such that blocks are generated at an average rate of one every ten minutes.

        \paragraph{Mining} 
        
When a miner creates a block, she is compensated for her efforts with Bitcoins. This compensation includes a per-transaction fee paid by the users whose transactions are included, as well as an amount of new Bitcoins that did not exist before. 

        \paragraph{Forks} 

Any miner may add a valid block to the chain by simply publishing it over an overlay network to all other miners. 
If multiple miners create blocks with the same preceding block, the chain is \emph{forked} into \emph{branches}, forming a tree. Other miners may subsequently add new valid blocks to any of these branch. When a miner tries to add a new block after an existing block, we say it \emph{mines on} the existing block. If this block is a leaf of a branch, we say he mines on the branch. 

To resolve forks, the protocol prescribes on which chain the miners should mine. 
The criterion is that the winning chain is the \emph{heaviest one}, that is, the one that required (in expectancy) the most mining power to generate. 
All miners add blocks to the heaviest chain of which they know, with random tie-breaking.\footnote{Choosing a longest branch at random is suggested in~\cite{eyal2014majority}. The operational client currently chooses the first branch it has heard of, making it more vulnerable
(see~\cite{eyal2014majority} for details).} 
The heaviest chain a node knows is the serialization of RSM inputs it knows, and hence describes the RSM's state. 
The formation of forks is undesirable, as they indicate that there is no globally-agreed RSM state. 

Branches and blocks outside the main chain are called pruned.\footnote{Often confusingly referred to as orphans in informal discussions, despite their having a parent in the block tree.} 
Transactions in pruned blocks are ignored. 
They can be placed in the main chain at any later time, unless a contradicting transaction (that spends the same outputs) was placed there in the meantime. 

Block dissemination over the Bitcoin overlay network takes seconds, whereas the average mining interval is ten minutes. Therefore, accidental bifurcation is rare.  It occurs on average once about every~60 blocks~\cite{decker2013propagation}.

    \section{Bitcoin-NG} \label{sec:protocol} 


\bitcoinNG\ is a blockchain protocol that serializes transactions, much like Bitcoin, but allows for better latency and bandwidth without sacrificing other properties. 

The protocol divides time into epochs. 
In each epoch, a single leader is in charge of serializing state machine transitions. 
To facilitate
state propagation, leaders generate blocks.
The protocol introduces two types of blocks: \emph{key blocks} for leader election and \emph{microblocks} that contain the ledger entries. 
Each block has a header that contains, among other fields, the unique reference of its predecessor, namely a cryptographic hash of the predecessor header. 

The security of the protocol derives from its incentive compatibility, motivating the participants to follow the rules. We detail the operation of the protocol in this section and explain its incentive system in Section~\ref{sec:security}. 

        \subsection{Key Blocks and Leader Election} 

Key blocks are used to choose a leader. 
Like a Bitcoin block, a key block contains the reference to the previous block, the current GMT time, a coinbase transaction to pay out the reward, a target value, and a nonce field containing arbitrary bits. 
For a key block to be valid, the cryptographic hash of its header must be smaller than the target value. 
Unlike Bitcoin, a key block contains a public key that will be used in the subsequent microblocks. 
 
As in Bitcoin, for a miner to generate a key block it must iterate through nonce values until the crypto-puzzle condition is met. 
As a result, the interval between consecutive key blocks is exponentially distributed.
To maintain a set average rate, the difficulty is adjusted by deterministically changing the target value based on the GMT time in the key block headers. 

In case of a fork, the chain is defined to be the one which represents the most work done, aggregated over all key blocks, with random tie breaking. 

        \subsection{Microblocks} 

Once a node generates a key block it becomes the leader. 
As a leader, the node is allowed to generate microblocks at a set rate smaller than a predefined maximum. 
The maximum rate is deterministic, and can be much higher than the average interval between key blocks. 
The size of microblocks is bounded by a predefined maximum. 
Specifically, if the timestamp of a microblock is in the future, or if its difference with its predecessor's timestamp is smaller than the minimum, then the microblock is invalid. 
This bound prohibits a leader (malicious, greedy, or broken) from swamping the system with microblocks. 

\begin{figure}[t] 
\centering 
\includegraphics[width=0.5\linewidth]{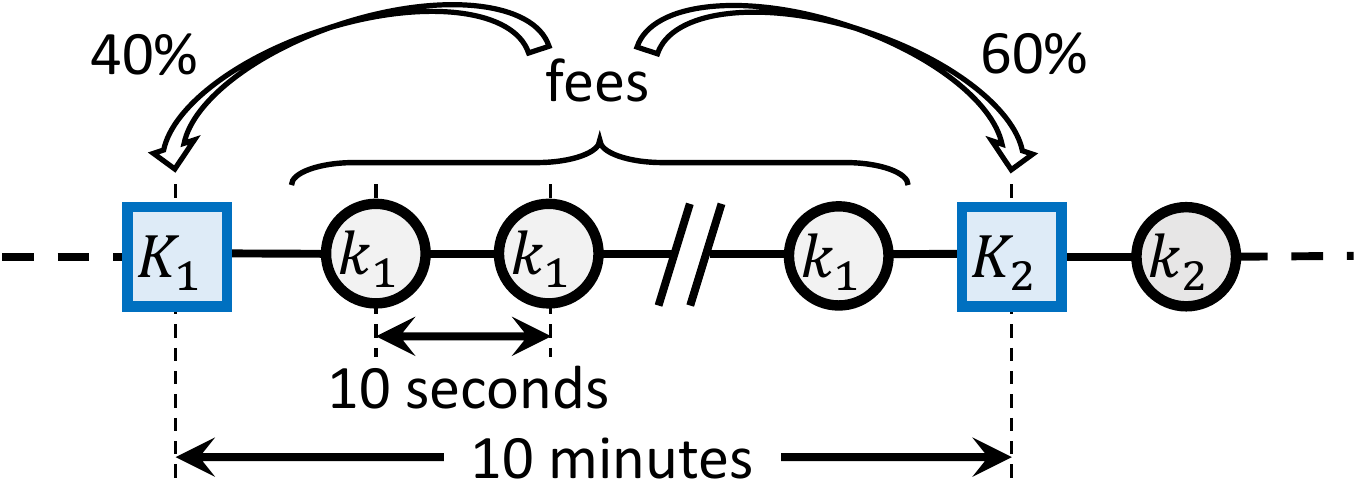} 
\caption[.]{\protect
Structure of the \bitcoinNG\ chain. Microblocks (circles) are signed with the private key matching the public key in the last key block (squares). Fee is distributed~$40\%$ to the leader and $60\%$ to the next one. 
}
\label{fig:ngChain} 
\end{figure}

A microblock contains ledger entries and a header. 
The header contains the reference to the previous block, the current GMT time, a cryptographic hash of its ledger entries, and a cryptographic signature of the header. 
The signature uses the private key that matches the public key in the latest key block in the chain. 
For a microblock to be valid, all its entries must be valid according to the specification of the state machine, and the signature has to be valid.
Figure~\ref{fig:ngChain} illustrates the structure. 

Note that microblocks do not affect the weight of the chain, as they do not contain proof of work. This is critical for maintaining the incentives aligned, as we explain in Section~\ref{sec:security}. 

        \subsection{Confirmation Time} 

\begin{figure}[t] 
\centering 
\includegraphics[width=0.5\linewidth]{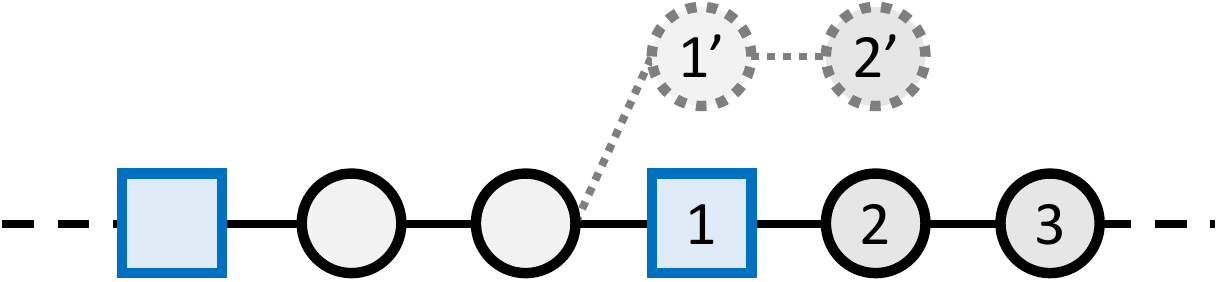}
\caption[.]{\protect
When microblocks are frequent, short forks occur on almost every leader switch. 
}
\label{fig:microBlockFork} 
\end{figure}

When a miner generates a key block, he may not have heard of all microblocks generated by the previous leader. 
If microblock generation is frequent, this can be the common case on leader switching. 
The result is a short microblock fork, as illustrated in Figure~\ref{fig:microBlockFork}. 
This fork is observed by any node that receives the to-be-pruned microblock (blocks 1' and~2' in the figure) before the new key block (block~1 in the figure). 
It is resolved once the key block propagates to that node. 
Therefore, a user that sees a microblock should wait for the propagation time of the network before considering it in the chain, to make sure it is not pruned by a new key block. 

        \subsection{Remuneration} 

To motivate mining, a leader is compensated for her efforts by the protocol. 
Remuneration is comprised of two parts. 
First, each key block entitles its generator a set amount. 
Second, each ledger entry carries a fee. This fee is split by the leader that places this entry in a microblock, and the subsequent leader that generates the next key block. 
Specifically, the current leader earns $40\%$ of the fee, and the subsequent leader earns $60\%$ of the fee, as illustrated in Figure~\ref{fig:ngChain}. The choice of this distribution is explained in Section~\ref{sec:security}. 

In practice, the remuneration is implemented by having each key block contain a single coinbase transaction that mints new coins and deposits the funds to the current and previous leaders. 
As in Bitcoin, this transaction can only be spent after a maturity period of~100 blocks, to avoid non-mergeable transactions following a fork. 

        \subsection{Microblock Fork Prevention} 

Since microblocks do not require mining, they can cheaply and quickly be generated by the leader, allowing it to split the brain of the system, publishing different replicated-state-machine states to different machines. 
This allows for double spending attacks, where different nodes believe the same coins were spent with different transactions. 

To demotivate such behavior, we use a dedicated ledger entry that invalidates the revenue of fraudulent leaders. 
Such entries have been used in different contexts~\cite{eyal2015cache, back2014sidechains, buterin2015slasher}. 
In \bitcoinNG, the entry is called a \emph{poison transaction}, and it contains the header of the first block in the pruned branch as a proof of fraud. 
The poison transaction has to be placed after the subsequent key block, and before the revenue is spent by the malicious leader. 
Besides invalidating the compensation sent to the leader that generated the fork, a poison transaction grants the current leader a fraction of that compensation, e.g.,~$5\%$. 
The choice of this value is explained in Section~\ref{sec:security}. 

Only one poison transaction can be placed per cheater, even if the cheater creates many forks. 
The cheater's revenue funds not relayed to the poisoner are lost.

    \section{Security Analysis} \label{sec:security} 

        \subsection{Incentives} 

Miners with capacity smaller than~$1/4$ of the total network are incentivized to follow the protocol. 
Specifically, miners are motivated to (1) include transactions in their microblocks, (2) extend the heaviest chain, and (3) extend the longest chain. Unlike Bitcoin, the latter two points are not identical. 


            \paragraph*{Heaviest Chain Extension} 

The motivation for extending the heaviest chain is the same as in Bitcoin. 
Since the majority will extend the heaviest chain, it will remain the main chain with high probability. 
A minority choosing to mine on another branch will not catch up, therefore it will mine on the main chain to ensure its revenues.\footnote{A majority may arbitrarily switch to any branch and win~\cite{kroll2013economics}.} 
\bitcoinNG\ therefore achieves the Nakamoto consensus Termination and Agreement under the postulate that Bitcoin does~\cite{miller2009model}. 

Microblocks carry no weight, not even as a secondary index. If they did, it would increase the system's vulnerability to selfish mining~\cite{eyal2013broken,nayak2015stubborn,sapirshtein2015optimal}. 
In selfish mining, an attacker withholds blocks it has mined and publishes them judiciously to obtain superior presence in the main chain. 
If microblocks had carried weight, an attacker could keep secret microblocks and gain advantage by mining on microblocks unpublished to anyone else. 

We conclude that \bitcoinNG\ does not introduce a new vulnerability to selfish mining strategies, and so \bitcoinNG\ is resilient to selfish mining against attackers with less than~$1/4$ of the mining power. 
\bitcoinNG\ therefore achieves the Nakamoto consensus validity under the postulate that Bitcoin does. 

            \paragraph*{Transaction Inclusion}

A leader earns~$40\%$ of a transaction's revenue by placing it in a microblock. 
However, he could potentially improve his revenue by secretly trying to earn~$100\%$ of the fee. 
To do so, first, the leader creates a microblock with the transaction, but does not publish it. 
Then, he tries to mine on top of this secret microblock, while other miners mine on older microblocks. 
If the leader succeeds in mining the subsequent key block, he obtains~$100\%$ of the transaction fees. 
Otherwise, he waits until the transaction is placed in a microblock by another miner and tries to mine on top of it. 

Denote by $r_\text{leader}$ the revenue of the leader from a transaction, leaving $(1 - r_\text{leader})$ for the next miner. 
In \bitcoinNG\ we have $r_\text{leader} = 40\%$. 
The value of $r_\text{leader}$ has to be such that the average revenue of a miner trying the above is smaller than his revenue placing the transaction in a public microblock as it should: 
\begin{equation*} 
    \overbrace{
        \alpha \times 100\%
    }^{\text{Win } 100\%}
+ 
    \overbrace{
        (1 - \alpha) \times \alpha \times (100\% - r_\text{leader})  
    }^{\text{Lose } 100\% \text{, but mine after txn}}
< 
    r_\text{leader}
\,\, , 
\end{equation*} 
therefore $r_\text{leader} > 1 - \frac{1 - \alpha}{1 + \alpha - \alpha^2}$. 
Assuming the power of an attacker is bounded by~$1/4$ of the mining power, we obtain $r_\text{leader} > 37\%$, hence $r_\text{leader} = 40\%$ is within range. 

            \paragraph*{Longest Chain Extension}

To increase his revenue from a transaction, a miner could avoid the transaction's microblock and mine on a previous block. 
Then he would place the transaction in its own microblock and try mining the subsequent key block. 
His revenue in this case must be smaller than his revenue by mining on the transaction's microblock as prescribed: 
\begin{equation*} 
    \overbrace{
        r_\text{leader} \vphantom{(100\%)}
    }^{\substack{\text{Place in}\\\text{microblock}}}
+ 
    \overbrace{
        \alpha (100\% - r_\text{leader})
    }^{\substack{\text{Mine next}\\\text{key block}}}
< 
    \overbrace{
        100\% - r_\text{leader} \vphantom{(100\%)}
    }^{\substack{\text{Mine on existing}\\\text{microblock}}}
\,\, , 
\end{equation*} 

therefore $r_\text{leader} < \frac{1 - \alpha}{2 - \alpha}$. 
Assuming the power of an attacker is bounded by~$1/4$ of the mining power, we obtain $r_\text{leader} < 43\%$, hence $r_\text{leader} = 40\%$ is within range. 

%

            \paragraph*{Optimal Network Assumption} 

One may assume a zero latency network where an attacker cannot rush messages~--- receive a message and send its own such that other nodes receive the attacker's message before the original one. 
Under such assumptions, Bitcoin is believed to be secure against selfish mining attackers of size up to almost~$1/3$~\cite{sapirshtein2015optimal}. 
However, for \bitcoinNG, 
due to the conditions above, 
we obtain $r_\text{leader} > 45\%$ and $r_\text{leader} < 40\%$, 
leaving no intersection. 
Under such optimal network assumptions, Bitcoin's blockchain is therefore more resilient than \bitcoinNG. 

            \paragraph*{Bypassing Fee Distribution}
            
We note that a user can circumvent the~$40-60\%$ transaction fee distribution by paying no transaction fee, and instead paying the current leader directly, using the coinbase address of the leader's key block. 
However, a user does not gain a significant advantage by doing so. 
As we have seen above, paying only the current leader increases the direct motivation of the current leader to place the transaction in a microblock, but reduces the motivation of future miners to mine on this microblock. 
Moreover, if the leader does not include the transaction before the end of its epoch, subsequent leaders will have no motivation to place the transaction. 

Other motives for fee manipulation such as paying a large fee to encourage miners to choose a certain branch after a fork apply to Bitcoin as well as \bitcoinNG, and are outside the scope of this work. 



        \subsection{Other concerns} 

            \paragraph{Wallet Security} 

The possibility of placing a poison transaction allows an attacker that obtains a leader's private key to revoke his revenue retroactively and earn a small amount. 
However, such an attacker is better off trying to steal the full leader's revenue when it becomes available, therefore the introduction of the poison transaction does not add a significant vulnerability. 

            \paragraph{Censorship Resistance} 

A central goal of Bitcoin is to prevent a malicious discriminating miner from dropping a user's transactions. 

First, we note that a leader's absolute power is limited to his epoch of leadership. 
A malicious leader can perform a DoS attack by placing no transactions in microblocks. 
Similarly, a benign leader that crashes during his epoch of leadership will publish no microblocks. 
Their influence ends once the next leader publishes his key block. 
The impact of such behaviors is therefore similar to that in Bitcoin, where nodes may mine empty blocks, but rarely do. 

Assuming an honest majority and no backlog, a user will have her transaction placed in the first block generated by an honest miner. 
At least $3/4$ of the blocks are generated by honest miners, therefore the user will have to wait for $4/3$ blocks on average, or~$13.33$ minutes. 
The frequent microblocks of \bitcoinNG\ do not improve censorship resistance. 
Key block intervals can be set to a rate that would reduce censorship to the minimum allowed by the network without incurring prohibitive deterioration of other metrics. 

            \paragraph{Resilience to Mining Power Variation} 

Following Bitcoin's success, hundreds of alternative currencies were created~\cite{wiki2013list}, most with Bitcoin's exact blockchain structure, and many with the same proof-of-work mechanism. 
To maintain a stable rate of blocks, different instances of the Blockchain tune their proof of work difficulty at different rates: 
Bitcoin once every~2016 blocks~-- about~2 weeks, Litecoin~\cite{litecoin2013site} every~2016 blocks (produced at a higher rate)~-- about 3.5 days, and Ethereum~\cite{ethereum2015white} on every block~-- about~12 seconds. 
However, whichever adjustment rate is chosen, these protocols are all sensitive to sudden mining power drops. 
Such drops happen when miners are incentivized to stop mining due to a drop in the currency's exchange rate, or to mine for a different currency that becomes more profitable due to a change in mining difficulty or exchange rate of either currency. 
Such changes are especially problematic for small alt-coins.
When their value rises, they observe a rapid rise in mining power as miners flock to reap easy revenues. 
Then, once the difficulty rises, the miners move on to mine on more profitable alt-coins and the mining power of the former drops. 
Now, since the difficulty is high, the remaining miners will need a longer time to generate the next block, potentially orders of magnitude longer. 

In \bitcoinNG, difficulty adjustments can create a similar problem, however it only affects key blocks. 
Microblocks are generated at the same constant rate. 
As a consequence, in case of a sudden mining power drop, \bitcoinNG's censorship resistance is reduced, as key blocks are generated infrequently. 
If a malicious miner becomes leader, it will generate microblocks until an honest leader finds a key block. 
Nevertheless, transaction processing continues at the same rate, in microblocks. 
Additionally, even until the difficulty is tuned to a correct value, the ratio of time during which malicious miners are leaders remains proportional to their mining power. 

            \paragraph{Forks} 

When issuing microblocks at a high frequency, \bitcoinNG\ observes a fork almost on every key block generation, as the previous leader keeps generating microblocks until it receives the key block (Figure~\ref{fig:microBlockFork}). 
These forks are resolved quickly~--- once the new key block arrives at a node, it switches to the new leader. 
In comparison, when running Bitcoin at such high frequency, forks are only resolved by the heaviest chain extension rule, and since different miners may mine on different branches, branches remain extant for a longer time compared to \bitcoinNG. 

\begin{figure}[t] 
\centering 
\includegraphics[width=0.7\linewidth]{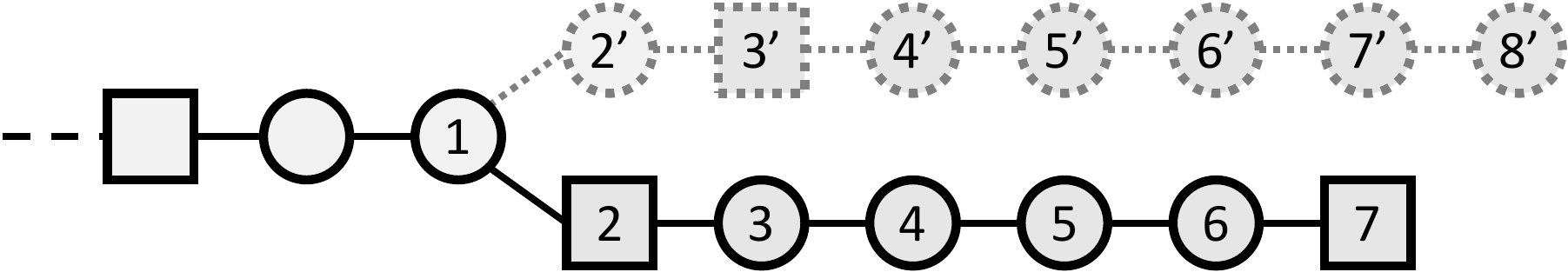}
\caption[.]{\protect
Key block fork. Blocks~2 and~3' have the same chain weight, and the fork is not resolved until key block~7 is generated. 
}
\label{fig:keyBlockFork} 
\end{figure}

However, \bitcoinNG\ may experience key block forks, where more than one key blocks is generated after the same prefix of key blocks, as shown in Figure~\ref{fig:keyBlockFork}. 
This rarely happens, due to low frequency and quick propagation of the small key blocks. 
However, the duration of the fork in this case may be very long, because it is only resolved on the next key block generation. 
The result is therefore infrequent, but long, key block forks. 

Although such long forks are undesirable, they are not dangerous. The knowledge of the fork is propagated through the network, and once it reaches the nodes, they are aware of the undetermined state. 
All transactions that appear only on one branch are therefore uncertain, until one branch gains a lead. 


    \section{Metrics} \label{sec:metrics}

We now detail the metrics we shall use to evaluate Bitcoin and \bitcoinNG. These metrics are designed to evaluate the unique properties of the Nakamoto consensus. 

%

        \paragraph{Consensus Delay} 

Intuitively, \emph{consensus delay} is  the time it takes for a system to reach agreement. 
We start by defining, for a specific execution and time, how long back nodes have to look to find a point where they agree on the state. 

\begin{figure}[t] 
\centering 
\includegraphics[width=0.5\linewidth]{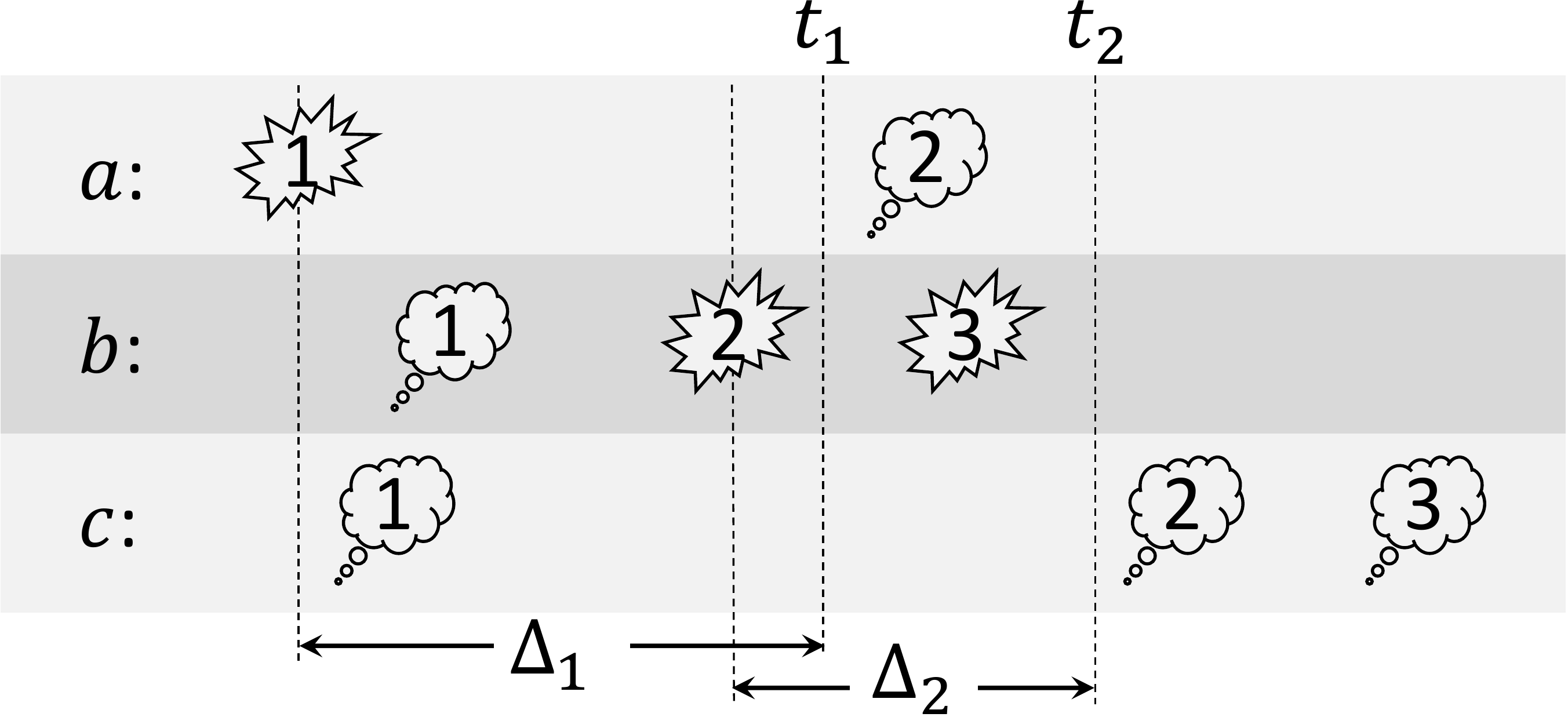}
\caption[.]{\protect
Point-consensus delay example with three Bitcoin nodes~$a$,~$b$, and~$c$ that generate blocks at heights 1, 2, and 3 (explosions) and learn that these blocks are in the main chain (clouds). Intervals~$\Delta_1$ and~$\Delta_2$ are the~$2/3$-point consensus delays at times~$t_1$ and~$t_2$, respectively. 
}
\label{fig:pointConsensusDelay} 
\end{figure}

In a specific execution of an algorithm, given a time~$t$ and a ratio $0 < \varepsilon \leq 1$, the \emph{$\varepsilon$ point consensus delay} is the smallest time difference~$\Delta$ such that at least $\varepsilon \cdot |\nodes|$ of the nodes at time~$t$ report the same state machine transition prefix up to time~$t - \Delta$. 
An example for the Bitcoin protocol is illustrated in Figure~\ref{fig:pointConsensusDelay}. 

The consensus delay is the best point-consensus-delay the system achieves for a certain fraction of the time, on average. 
More formally, the \emph{$(\varepsilon, \delta)$ consensus delay} of a system is the $\delta$-percentile $\varepsilon$-point-consensus-delay. For example, if during at least~$90\%$ of the time, at least~$50\%$ of the nodes agree on the state of the state machine~10 seconds ago, then the~$(50\%, 90\%)$-consensus delay is~10 seconds. 

        \paragraph{Fairness} 

We calculate two ratios: 
(1) the ratio of transitions not coming from the largest miner with respect to all transitions, and 
(2) the ratio of mining power not owned by the largest miner with respect to all mining power. 
We call the ratio of these ratios the \emph{fairness}. 

Optimally the fairness is~1.0: The largest miner and the non-largest miners' representation in the transitions set should be the same as their respective mining powers. 

        \paragraph{Mining Power Utilization} 

The security of a proof-of-work system derives from the mining power used to secure it; that is, the mining power an attacker has to outrun in order to obtain disproportionate control. 
The \emph{mining power utilization} is the ratio between the mining power that secures the system and the total mining power. 
Mining power wasted on work that does not appear on the blockchain accounts for the difference. 

        \paragraph{Subjective Time to Prune} 

Due to the probabilistic nature of the Nakamoto consensus, a node may learn of a state machine transition and subsequently learn that this transition has not occurred~-- that it was pruned from history. This is the case with pruned branches in Bitcoin. 

\begin{figure}[t] 
\centering 
\includegraphics[width=0.5\linewidth]{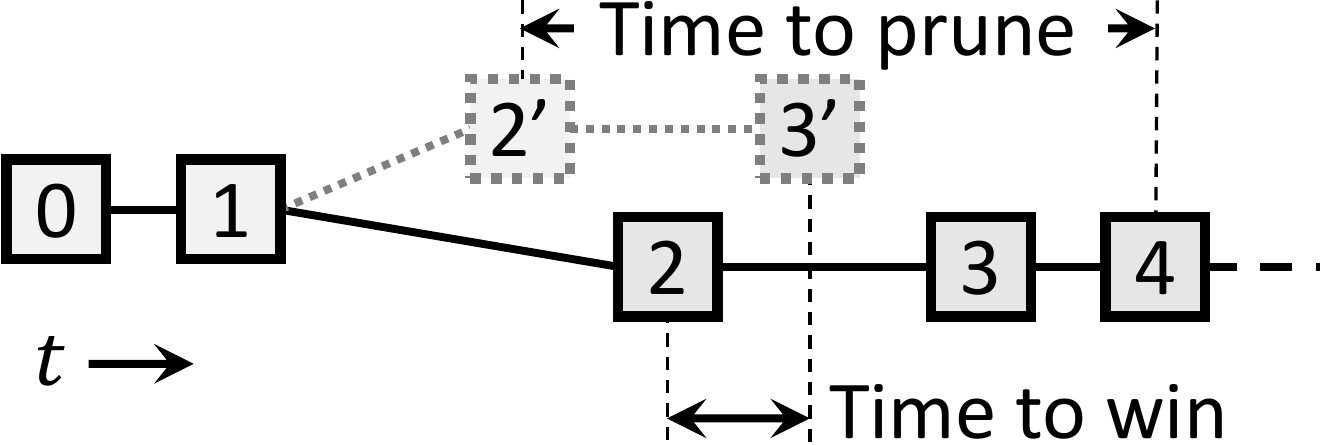}
\caption[.]{\protect
The time-to-prune and time-to-win metrics. 
}
\label{fig:timeTo} 
\end{figure}

The $\delta$ time to prune is the $\delta$-percentile of the difference between the time a node learns about such a transition and the time it learns that this transition has not occurred. 
This implies what time a user has to wait to be confident a transition has occurred. 
Figure~\ref{fig:timeTo} illustrates an example for the Bitcoin protocol.

        \paragraph{Time to Win} 

The $\delta$ time to win is the $\delta$ percentile of the difference between the first time a node believes a never-to-be-pruned-transition has occurred and the last time a (different) node disagrees, believing an alternative transition has occurred. It is zero if the latter time is earlier. 
Figure~\ref{fig:timeTo} illustrates an example for the Bitcoin protocol.

    \section{Experimental Setup} \label{sec:setup} 

We evaluate Bitcoin and \bitcoinNG\ with~1000-node experiments on an emulated network. 

        \paragraph{Implementation} 

For Bitcoin we run the standard client (release~0.10.0), hereinafter \emph{Bitcoin}, with minimal instrumentation to log sufficient information. 

We implemented all \bitcoinNG\ elements that are significant for a performance analysis in the absence of an adversary, by modifying the standard Bitcoin client (release~0.10.0). 
We did not implement the fee distribution and the microblock signature check. 
Both elements have negligible implication on performance~--- fee distribution requires about one fixed point operation per transaction and signature checking adds several milliseconds per microblock. 

        \paragraph{Simulated Mining}

The time it takes a miner to find a solution follows a geometric probability distribution, which can be approximated as an exponential distribution due to the improbability of a success in each guess and the rate of guessing.

In our experiments we replace the proof of work mechanism with a scheduler that triggers block generation at different miners with exponentially distributed intervals. This is implemented using the regression-test mode of the standard Bitcoin client~\cite{bitcoin2015source} and an in-situ controller. In regression-test mode, the client skips the block difficulty validation and accepts blocks with any difficulty. It also accepts commands to instantly generate a block, which the scheduler uses.

        \paragraph{Mining Power} 

\begin{figure}[t] 
\centering 
\noindent \includegraphics[width=0.8\linewidth]{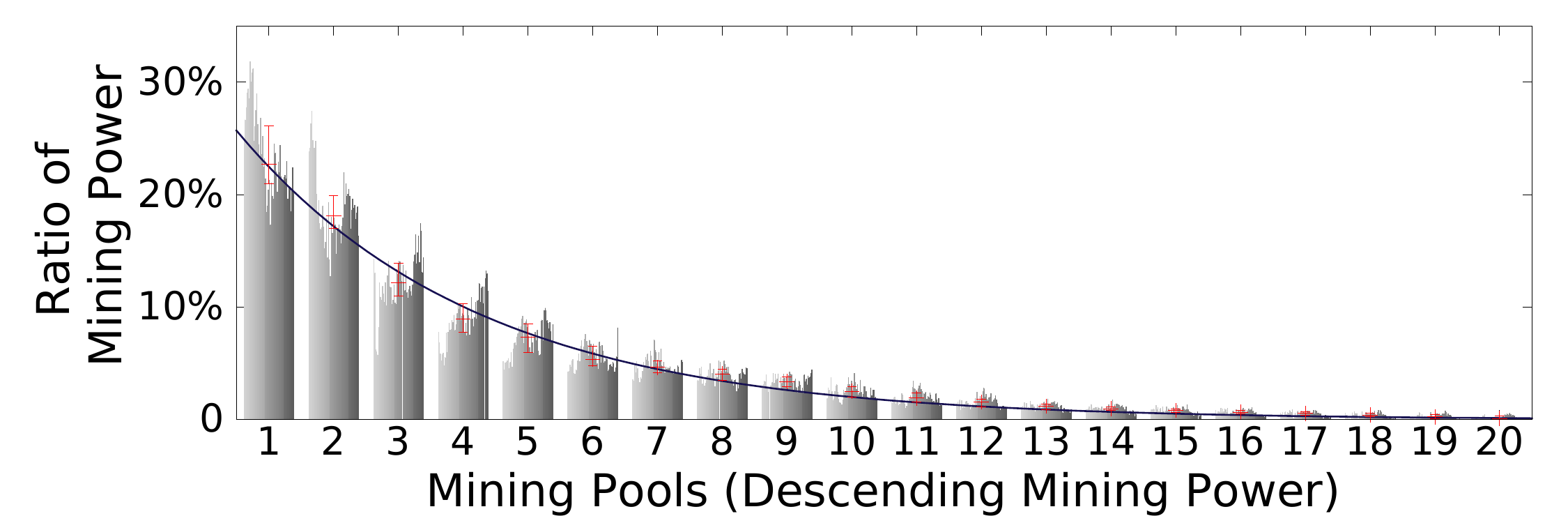}
\caption[.]{\protect
Error bars represent the 75th, 50th and 25th percentiles of the corresponding batch.
}
\label{fig:poolRatio} 
\end{figure} 

The probability of mining a block is proportional on average to the mining power used for solving the cryptopuzzle. 
Since blocks are generated at average set intervals and the total amount of mining power is large, the interval between block generation events of a small miner is extremely large. 
A single home miner using dedicated hardware is unlikely to mine a block for years~\cite{swanson2013calculator}. 

Consequently, mining power tends to centralize in the form of industrial mining and open mining pools. 
Industrial miners are companies that operate large-scale mining facilities. 
Smaller miners that run private mining rigs typically join forces and form mining \emph{pools}. All members of a pool work together to mine each block, and share their revenues when one of them successfully mines a block. 

To reflect in our setup the varying power of miners, we examined the power distribution in Bitcoin mining entities. 
The information we require for the analysis, the identity of the entities generating each block, is voluntarily provided by miners. 
We used a public API~\cite{blockTrailAPI} to gather this information for the year ending on August 31, 2015. 
We note that about $9\%$ of the blocks are unidentified. 
We considered each such block as generated by a different individual miner. 

For each week of the year, we calculate the \emph{weekly mining power} of each entity, and assign rank~1 to the largest weekly mining power, rank~2 for the second largest, and so on. 
Figure~\ref{fig:poolRatio} shows the weekly mining power of each entity by rank up to~20. 
Bars of the same shade at different ranks show the distribution of a specific week. 
Each batch of bars represents the collection of ratios for the $n\textsuperscript{th}$ highest block generating pool. 
We note that the ranks of different entities is not preserved throughout the weeks. 
The y-axis represents the weekly ratio of blocks generated by a pool. 

To model the size distribution of mining entities, we approximate it with an exponential distribution with an exponent of $-0.27$. 
It yields a~$0.99$ coefficient of determination compared with the medians of each rank. 


        \paragraph{Network}

The structure of Bitcoin's overlay network is complicated, and much of it is intentionally hidden to preserve Bitcoin's security against denial of service (DoS) and to maintain participants' privacy~(see~\cite{Heilman2015eclipse,miller2015topology} for details on the peer-to-peer network). 
Nodes do not reveal their neighbors, only a superset that includes nodes they have heard of. 
Many of the nodes are hidden behind firewalls making it difficult to even estimate the full size of the network. 
The latency among nodes is unknown. 
Moreover, for many of the metrics that we measure, a critical question is the time it takes between the mining of a block by some miner and the time it is being mined on by another miner. 
For this to happen, the block not only has to be propagated and verified by the second miner, but that second miner must also propagate the details to its mining hardware. 
In the case of mining pools with many distant worker miners, this may incur a non-negligible delay. 

Lacking an existing model of the system, we construct a random network by connecting each node to at least~5 other nodes, chosen uniformly at random. 
We measured the latency to all visible Bitcoin nodes from a single vantage point on~April 7th, 2015, and created a latency histogram. We then set the latency among each pair of nodes in the experiments based on this histogram. The bandwidth is set to about 100kbit/sec among each pair of nodes. 

\begin{figure}[t]
\centering
\includegraphics[width=0.6\linewidth]{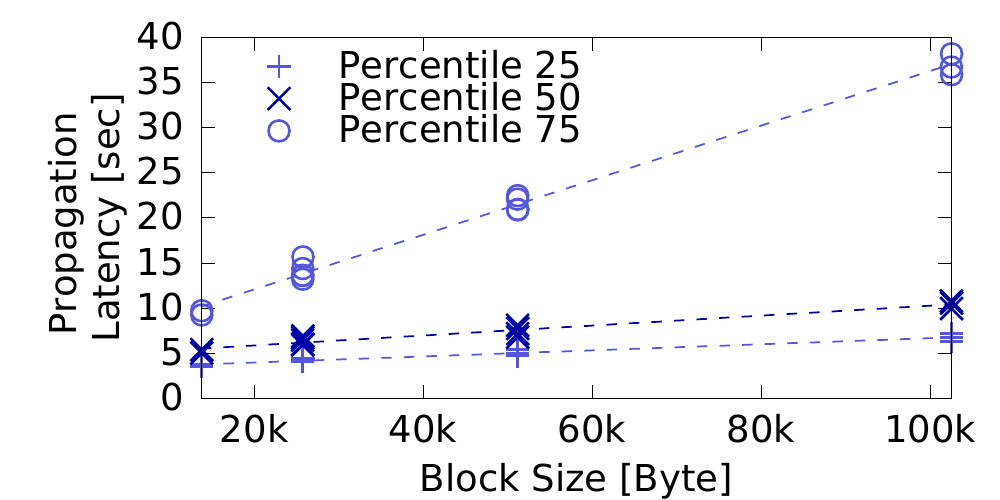}
\caption[.]{\protect 
In our system, block propagation time grows linearly with block size. 
This qualitatively matches the linear relation observed in measurements of the operational Bitcoin network~\cite{decker2013propagation}. 
}
\label{fig:propagation}
\end{figure}

To verify the validity of our setup and topology, we compare Bitcoin's propagation properties in our setup and in the operational system. 
We perform experiments with different block sizes while changing the block frequency so that the transaction-per-second load is constant. 
Figure~\ref{fig:propagation} shows a linear relation between the block size and the propagation time, similar to the linear relation measured in the Bitcoin operational network by Decker and Wattenhofer~\cite{decker2013propagation}. 

        \paragraph{No Transaction Propagation} 

The goal of this work is to optimize the consensus mechanism of the Blockchain. 
However, when generating blocks at high frequencies, the overhead of filling in the blocks by generating and propagating transactions becomes a dominant factor with Bitcoin's current implementation. 
This is not an inherent property of Bitcoin's protocol, or of a Blockchain protocol in general. 
In order to reduce the noise caused by the transaction generation and propagation mechanism, we reduce transaction handling to the minimum. 
Before starting an experiment, we initialize the blockchain with artificial transactions and top up the mempools (the data structure storing yet to be serialized transactions) of all nodes with the same set of independent transactions that can be serialized in arbitrary order. 
The transactions are of identical size; the operational Bitcoin system as of today, at~1MB blocks every~10 minutes, has a bandwidth of~3.5 such transactions per second.

\begin{figure}[!t]
\vfil
\centering
\subfloat[Reducing latency]{
\includegraphics[width=0.45\linewidth]{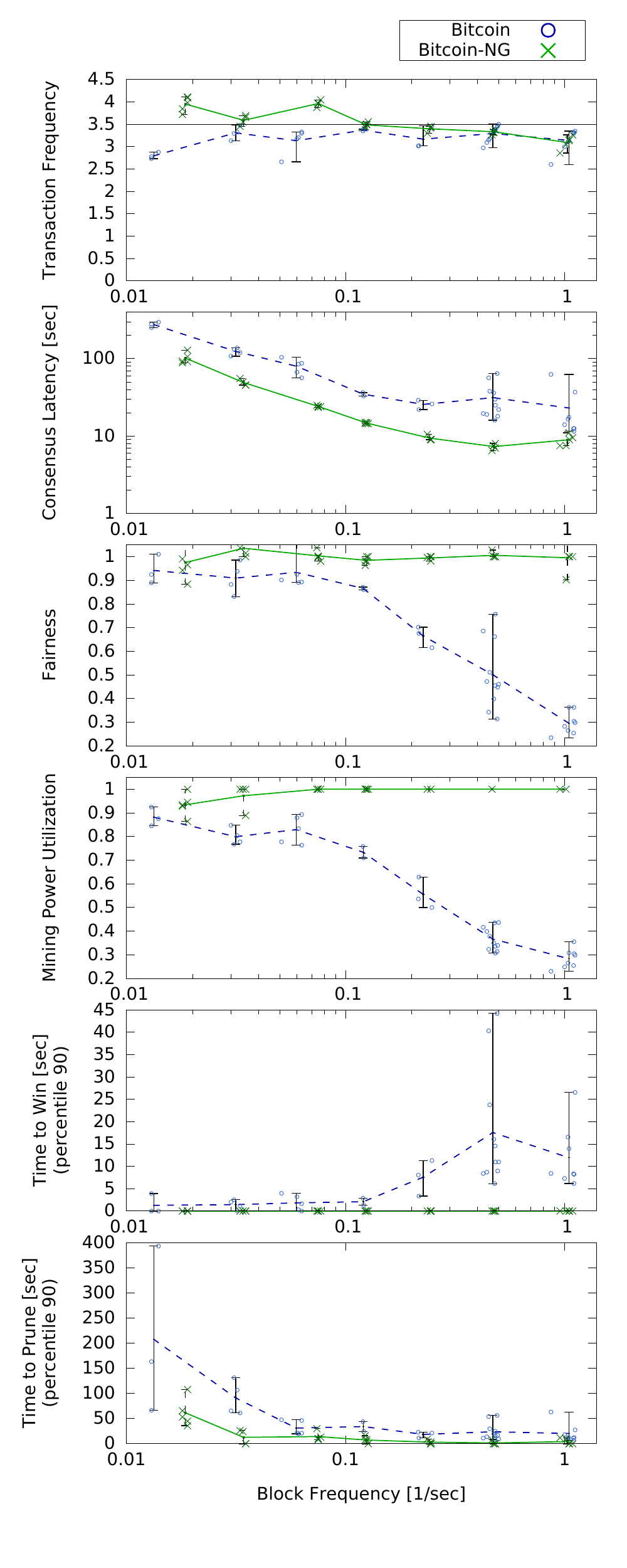}
}
\subfloat[Increasing throughput]{
\includegraphics[width=0.45\linewidth]{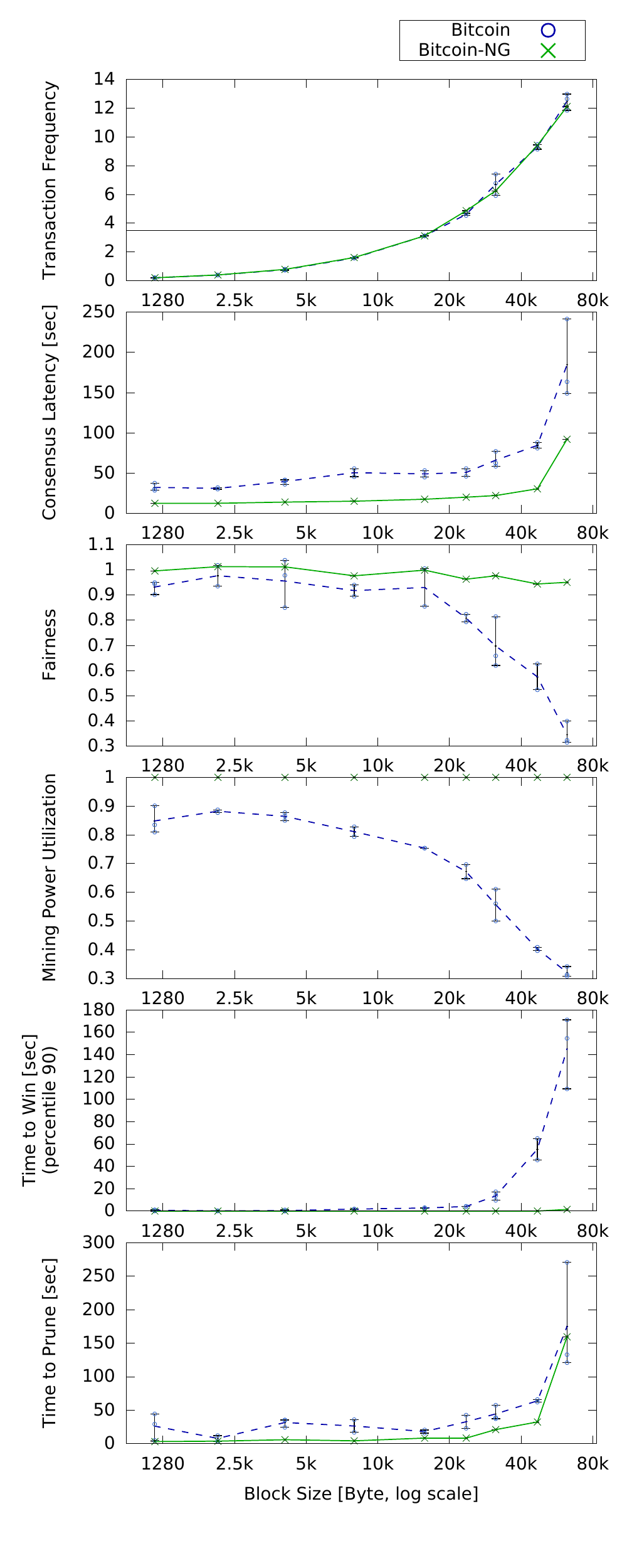}
}

\caption[.]{\protect
Experiment results 
}
\label{fig:results}
\end{figure}
\clearpage

    \section{Evaluation} \label{sec:evaluation}

We evaluate \bitcoinNG\ and compare it with Bitcoin in two sets of experiments, varying block frequency and block size. 

We observe that it is possible to improve Bitcoin's consensus delay and bandwidth by tuning its parameters, but its performance deteriorates dangerously on all security-related metrics. 
\bitcoinNG\ qualitatively outperforms Bitcoin, as it suffers no such deterioration, while enjoying superior performance in almost all metrics across the entire measured range. 

The bandwidth of \bitcoinNG\ is only limited by the processing speed of the individual nodes, as higher throughput does not introduce forks. The consensus delay is determined directly by the network propagation time, because in the common case all nodes agree on the main chain once they receive the latest key block. 


        \paragraph{Metrics} 

For each execution we run for~50-100 Bitcoin blocks or {\bitcoinNG} microblocks and, for each experiment, measure the metrics we introduced as follows:  

\begin{description} 
\itemsep0em

\item[Consensus delay] We take the $(90\%, 90\%)$-consensus delay based on block generation times. Point-con\-sensus-delay for Bitcoin is illustrated in Figure~\ref{fig:pointConsensusDelay}. 

\item[Fairness] We calculate the proportion of 
(1) the ratio of blocks in the main chain not generated by the largest miner with respect to all blocks in the main chain, and 
(2) the ratio of blocks not generated by the largest miner with respect to all generated blocks. 

\item[Mining power utilization] We calculate the proportion between the aggregate work of the main chain blocks and all blocks. 
In \bitcoinNG, difficulty is only accrued in key blocks, so microblock forks do not reduce mining power utilization. 

\item[Time to prune] For each node and for each branch, we measure the time it took for the node to prune this branch. This is the time between the receipt of the first branch block and the receipt of the main chain block that is longer than this branch (Figure~\ref{fig:timeTo}). We take the~\nth{90} percentile of all samples. 

\item[Time to win] We take the~\nth{90} percentile of the time from the generation of each main-chain block to the last time another miner generates a block that is not its descendant (Figure~\ref{fig:timeTo}). 

\end{description}

        \paragraph{Figures} 

We run multiple experiments with different parameters. The figures show the average value for each group of measurements with error bars marking the extreme values. The sampled values are shown as markers. 

        \subsection{Block Frequency} 

First, we run experiments targeted at improving the consensus delay. 
For Bitcoin, we vary the frequency of block generation by reducing the proof-of-work difficulty. 
For \bitcoinNG, keeping the key block generation at one every~100 seconds, we vary the frequency of microblock generation. 
For each frequency, we choose the block size (microblock size for \bitcoinNG) such that the payload throughput is identical to that of Bitcoin's operational system, that is, one~1MB block every~10 minutes. Figure~\ref{fig:results} shows the results.

We confirm that the bandwidth, measured as transaction frequency, is close to~3.5, the operational Bitcoin rate of for such transactions. 
In our experiments, Bitcoin's bandwidth is smaller than that of \bitcoinNG, giving Bitcoin a small advantage. 

As expected, a higher block frequency reduces Bitcoin's consensus latency as transactions are placed in the ledger at a higher frequency. 
Time to prune improves significantly as block frequency increases. 
Nevertheless, Bitcoin's frequent forks leave it with higher consensus latency and time to prune than \bitcoinNG. 
We note that although they can be made arbitrarily rare, key block forks do occur. 
Such key-block forks are only resolved once one branch has more key blocks than the others, resulting in a long time to prune if key block intervals are long. 

Bitcoin's mining power utilization drops quickly as frequency increases, tending towards~$1/4$, the size of the largest miner. At the extreme, block generation is so fast that by the time a miner learns of a block generated by another miner, that other miner has generated more blocks. Then, only the largest miner generates main chain blocks, and the other miners catch up. This also implies the deterioration of fairness, as forks are likely to be resolved by the largest miner extending its preferred branch. 
As miners struggle to catch up with the leading pack, slow miners mine on old blocks and the time to win metric increases. 

Since contention in \bitcoinNG\ is limited to key block generation, forks remain rare despite high frequencies of microblocks. 
Increasing the microblock frequency achieves consensus delay and time to prune reduction. All other metrics are unaffected and remain at the optimal level. 

In the low frequency experiments of \bitcoinNG, we observe a slight mining power utilization decrease and time to prune increase. 
This is an artifact of the experimental setup. 
We run the experiments over a set number of blocks, therefore these low contention experiments run for an extended period, enough to observe key block forks. 
In these experiments, the key block frequency of \bitcoinNG\ is similar to the block frequency of Bitcoin. 
Indeed, in low contention settings \bitcoinNG\ provides only a minor advantage over Bitcoin since its key blocks are small and propagate quickly. 


        \subsection{Block Size} 


To study bandwidth scalability, we run experiments with different block sizes. 
We use high frequencies to observe the systems' limits, setting Bitcoin's block frequency to 1/10sec and \bitcoinNG's microblock frequency to~1/10sec and key block frequency to~1/100sec. Figure~\ref{fig:results} shows the result. 

As required, the transaction frequency increases with block size; the horizontal line shows the operational Bitcoin rate. 

Large blocks take longer to verify and propagate. 
Therefore, although block frequency is constant, the time it takes for a miner to learn of a new block is longer, and so the chance for forks increases. 

These experiments demonstrate the expected tradeoff between bandwidth and latency. 
Consensus latency increases due to forks, as it takes longer to choose the main chain. The time to win also increases, as blocks take longer to catch up with the larger blocks, and so is time to prune due to the many forks. 

While this tradeoff may be acceptable, allowing for some hunt for a sweet spot on the tradeoff curve, the real problem pertains to security. 
The forks cause significant mining power loss, reaching about~$80\%$ at Bitcoin's bandwidth (though at a higher block frequency), making the system vulnerable to attackers that are much smaller. 

Even more detrimental is the reduction in fairness. 
Even a minor degradation in fairness is dangerous, since it provides incentives to miners to avoid losses by joining forces to enjoy the advantage of mining in a larger pool. 
This leads to centralization of the mining power, obviating Bitcoin's security properties. 

\bitcoinNG\ demonstrates qualitative improvement, suffering no significant degradation in the security-related metrics of fairness and mining power. At high bandwidth, however, the clients are approaching their capacity, making it hard for them to keep up and we observe degradation in consensus latency and time to prune. 

    \section{Related Work} \label{sec:related} 

    \paragraph{Model} 

As in Bitcoin~\cite{nakamoto2008bitcoin} and enhancements thereof~\cite{ethereum2015white,sompolinsky2015ghost, lewenberg2015inclusive}, the goal of \bitcoinNG\ is to implement an RSM in an open system.
The exact assumptions and guarantees are explored in different works~\cite{bonneau2015sok,miller2009model,garay2015backbone}. 
Our model is similar to those of Aspnes et al.~\cite{aspnes2003survey} and Garay et al.~\cite{garay2015backbone}, and our definition of the Nakamoto Consensus is similar to that of~\cite{garay2015backbone}. 
These are different from the model and goal of classical Byzantine fault tolerant RSMs. 
Those, by and large, (1) assume static or slow to change membership, allowing for quorum systems and reconfigurations thereof, and (2) do not guarantee fairness of representation of honest parties in the state machine transitions. 


The problem of leader election was apparently first formulated and solved in~1977 by Gerard LeLann~\cite{le1977distributed}. 
In 1982, Hector Garcia-Molina addressed the problem in a distributed system that admits failures~\cite{garcia1982elections}. 
Since then leader election has been extensively used to improve the performance of distributed systems (e.g.,~\cite{dwork1988consensus,moraru2012egalitarian}). 
In these classical consensus protocols, the leader's role is to propose decisions that have to be confirmed by a quorum. This can be compared to having a block of a leader (as defined here) buried in blockchain protocols. 

    \paragraph{GHOST} 

The GHOST protocol of Sompolinsky et al.~\cite{sompolinsky2015ghost} improves on Bitcoin's scalability by changing its chain selection rule. 
While in Bitcoin the chain with the most work (accumulated over all chain blocks, based on their proof-of-work) is the main chain, with GHOST, at a fork, a node chooses the side whose sub-tree contains more work (accumulated over all sub-tree blocks). 
The benefit is that the heaviest sub-tree choice takes into account proof of work that does not end up in the main chain. 
Thus, GHOST improves both fairness and the mining power utilization under high contention. 

However, in GHOST, blocks on pruned subtrees only affect the selection rule at the branch point. 
The \bitcoinNG\ protocol maintains a small fork rate at high bandwidth and throughput, allowing for better mining power utilization and fairness. 
%
Moreover, to use GHOST in an operational system, a challenge remains. 
In Bitcoin, at any given time, at least one node knows what the main chain is since it knows all of its blocks. 
In GHOST this is not the case, and it is possible that no single node has enough information to determine which is the main chain.
Appendix~\ref{app:ghostExample} provides an example.

One solution to finding the true main chain in GHOST is to propagate all blocks.
However, this exposes the system to denial of service attacks, as a malicious node can overwhelm the network with low difficulty blocks. 
There may be heuristics to avoid the security danger; we do not address this question, but did evaluate the system by implementing it, propagating all blocks. 
Under these conditions, GHOST performed worse than Bitcoin as the overhead of propagating all blocks outweighed the benefits of the chain selection rule. 
Future work may find a solution to GHOST's practical challenges, e.g.\ by propagating only block headers. 
Such a practical implementation of GHOST can be used to complement \bitcoinNG\ and allow for a higher frequency of key blocks. 

    \paragraph{Inclusive Blockchains} 

Lewenberg et al.~\cite{lewenberg2015inclusive} replace the blockchain structure with a directed acyclic graph. There still is a main chain, but its blocks may refer to pruned branches to include their transactions. 
Analysis demonstrates considerable improvement of fairness and mining power utilization. 
\bitcoinNG\ achieves optimal fairness and mining power utilization. 
Using \bitcoinNG\ with an inclusive blockchain to increase key block frequency may prove problematic: 
Decommissioned leaders could retroactively introduce transactions and have them included by the current leader. 
This could allow for DoS and double spending attacks. 

\paragraph{Faster Bitcoin}

Significant effort by Bitcoin's core developers is put into improving the performance of the Bitcoin client and technical aspects of its protocol. While this work can provide significant improvement and enable better scaling, it does not eliminate the inherent limitation that stems from the forks forming at high rates.

Stathakopoulou et al.\ suggest to reduce propagation delay in the Bitcoin network~\cite{stathakopoulou2015faster}.
However, their suggestions imply significant compromises on security.
First, they have nodes propagate transaction inventory before having the actual transactions; this allows an attacker to swamp the network at no cost by publishing transaction IDs for non-existent transactions. 
Second, they form a network by having nodes prefer connections with close neighbors~--- exactly the opposite of the current security-oriented algorithm.

Improving the efficiency of the client~\cite{andresen2015o1, pazmino2015dividing} can improve propagation time and reduce the collision window (time before $A$ hears $B$ found a block). 
However the improvement is limited~--- a processing speed increase of $x\%$ allows for block size increase of $x\%$ at the same fork rate. 
\bitcoinNG\ provides a qualitative improvement that removes the fork rate dependency on block size or rate. 

Corallo~\cite{corallo2013relay} has built a centralized fast relay for Bitcoin, parallel to the standard peer-to-peer network. It significantly improves network throughput and latency but increases centralized control and reduces fairness~--- miners outside the fast relay are at a disadvantage. 

    \paragraph{Off-chain solutions} 

An alternative to improving the bandwidth and latency of the blockchain is to perform transactions off the chain. 
This basic premise apparently originated in Hearn and Spilman's two-point channel protocol~\cite{hearn2015contracts}. 
The Lightning network~\cite{poon2013lightning} and a protocol by Decker and Wattenhofer~\cite{decker2015duplex} allow for extensive payment networks where transactions occur without trusted middlemen. 
These solutions use contracts to allow any party to place fraud proof on the main blockchain and deny revenue from a villain. 

These solutions may be suitable in various scenarios, but they do not address the problem of scaling a Nakamoto-consensus RSM. 
As an extreme example, in the benign case of failure of the nodes performing transactions over a channel, all their transactions are lost, as they were never stored in the blockchain. 

Another proposition for improving bandwidth and latency is that of separate chains, known as side chains. 
In side chains, transactions can move Bitcoin from one chain to another~\cite{back2014sidechains}. 
This allows for sharding of the workload: A subset of the Bitcoins are moved to their own chain, and can subsequently be managed at that chain. 
Each shard, running in its own chain, can use \bitcoinNG\ to enjoy its efficiency. 
This solution does not increase efficiency if sharding is not possible and transactions frequently involve multiple chains.

    \paragraph{Analysis} 

Given a cryptopuzzle difficulty and a topology, Sompolinsky et al.~\cite{sompolinsky2015secure} calculate upper and lower bounds for the growth rate of the Bitcoin main chain. 
This analysis can be translated to the expected forking frequency at different difficulty levels when there are exactly two miners. 
Our experiments target a larger number of miners, modeled according to Bitcoin's operational system, that tune difficulty arbitrarily to reach a target main chain extension rate. 

Miller and Jansen~\cite{miller2015shadow} describe a methodology for evaluating a large-scale Bitcoin blockchain system on a single machine using an event-driven simulator. To facilitate manageable experiment times, they replace time-consuming cryptographic operations with a delay of an appropriate length. 
In our experiments, we run the original operational client directly on the operating system, emulating only the network properties.

    \section{Conclusion} \label{sec:conclusion} 

As Bitcoin and related cryptocurrencies have become surprisingly popular, they have hit scalability limits. The technical debate to improve scalability has been hampered by a perceived inherent tradeoff between performance metrics and security goals of the system. Consequently, the discussions have become acrimonious, long-term solutions have seemed elusive, and the current sentiment has centered around short-term, incremental, compromise solutions. 

\bitcoinNG\ shows that it is possible to improve the scalability of blockchain protocols to the point where the network diameter limits consensus latency and the individual node processing power is the throughput bottleneck.
Such scaling is key in allowing for blockchain technology to fulfill its promise of implementing trustless consensus for a variety of applications from payments, through digital asset transactions, to smart contracts~--- at global scale. 

\paragraph{Acknowledgements} The authors thank
Ayush Dubey,
Gregory Maxwell,
Malte M\"oser, and
Weijia Song
for their comments on initial versions of this manuscript.


{\footnotesize \bibliographystyle{acm}
\bibliography{btc}}

\onecolumn
\appendix 

    \section{GHOST Propagation Example} \label{app:ghostExample}

\begin{figure*}[!h]
\centering
\subfloat[View of node 1]{
\includegraphics[width=0.3\linewidth]{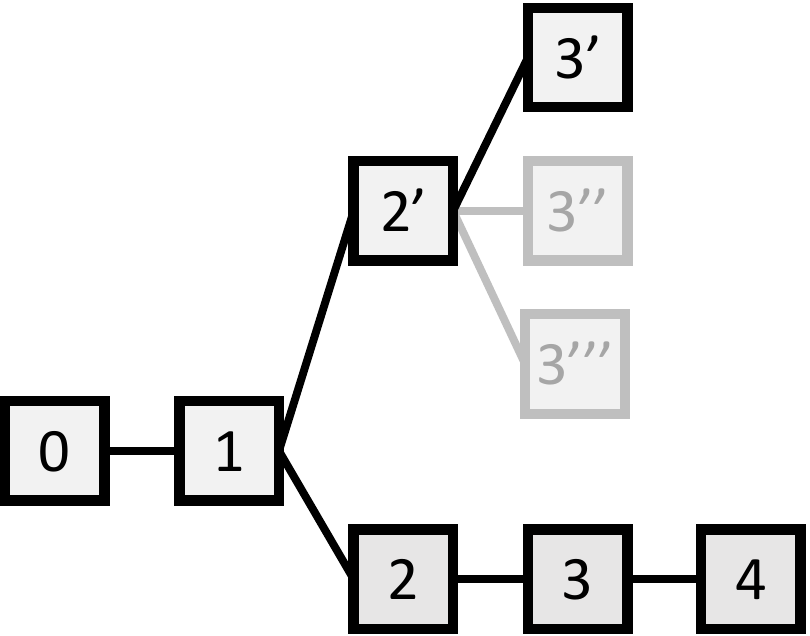}
\label{fig:ghostExample:node1}
}
\hfil
\subfloat[View of node 2]{
\includegraphics[width=0.3\linewidth]{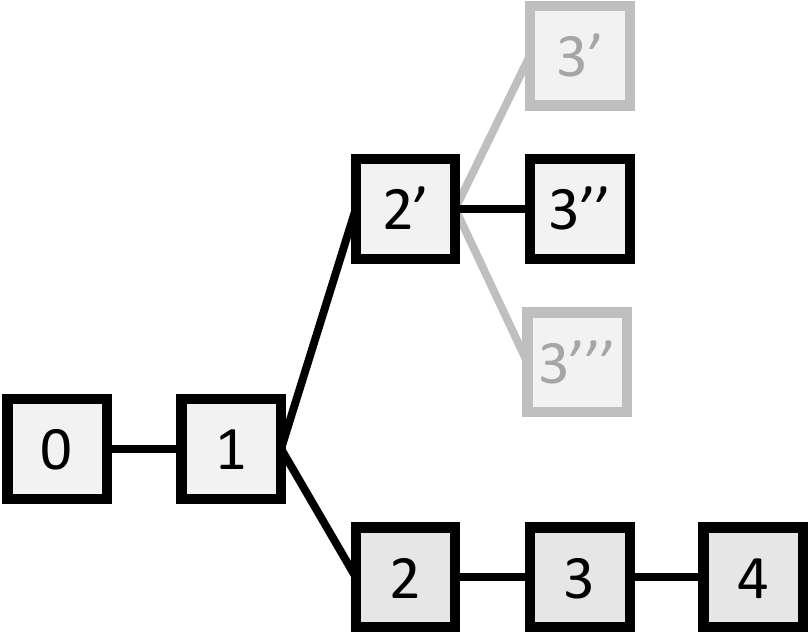}
\label{fig:ghostExample:node2}
}
\hfil
\subfloat[View of node 3]{
\includegraphics[width=0.3\linewidth]{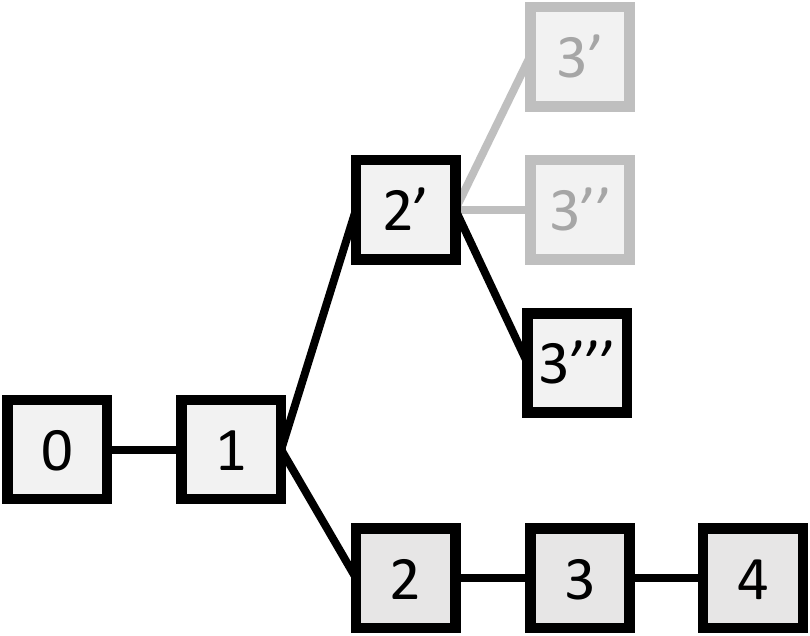}
\label{fig:ghostExample:node3}
}
\caption[.]{\protect
A partial view of the GHOST block tree by node~1 \subref{fig:ghostExample:node1}, node~2 \subref{fig:ghostExample:node2}, and~node~3 \subref{fig:ghostExample:node3} does not allow either of them to surmise which is the main chain.
}
\label{fig:ghostExample}
\end{figure*}

Figure~\ref{fig:ghostExample} illustrates an example of when no single GHOST protocol~\cite{sompolinsky2015ghost} node is aware of the main chain in the system. Consider three nodes, 1, 2, and 3, each of which is aware of only a subset of the blocks. Each node knows a chain with a length of height 4, and each knows of a branch of height~3 starting at a block~$2'$ and ending at either block~$3'$, $3''$,~or~$3'''$, as shown in Figures~\ref{fig:ghostExample:node1}, \ref{fig:ghostExample:node2}, and \ref{fig:ghostExample:node3}, respectively. 

    \section{Competition on a Key-Block Fork} \label{app:securityNotes}

We note that in case of a fork where two miners discover competing key blocks following the same key block (and after any number of subsequent microblocks) things become more complicated than they are in Bitcoin.
Here, each leader can publish transactions that pay a large fee to the subsequent miner in order to entice miners to choose his branch.
While this competition may introduce interesting dynamics beyond the scope of this work, we note that each branch may copy the transactions placed in the microblocks of the competing branch, and so even if an attacker is motivated to place significant fees due to external incentives, its competitor will copy those same transactions and remove the attacker's advantage.



\end{document}